\documentclass[aps,prb,reprint,superscriptaddress]{revtex4-1}

\usepackage{textcomp}
\usepackage{array}

\usepackage{graphicx}
\usepackage{float}
\usepackage{amsmath}
\usepackage{amssymb}
\usepackage{epstopdf}
\usepackage{bm}
\usepackage{natbib}

\def\0pi{0\!-\!\pi}

\begin{document}

\title{Ballistic superconductivity and tunable $\pi$-junctions in InSb quantum wells}

\author{Chung~Ting~Ke}\altaffiliation{Equal contributions}
\affiliation{QuTech and Kavli Institute of Nanoscience, Delft University of Technology, 2600 GA Delft, The Netherlands}

\author{Christian~M.~Moehle}\altaffiliation{Equal contributions}
\affiliation{QuTech and Kavli Institute of Nanoscience, Delft University of Technology, 2600 GA Delft, The Netherlands}

\author{Folkert~K.~de~Vries}
\affiliation{QuTech and Kavli Institute of Nanoscience, Delft University of Technology, 2600 GA Delft, The Netherlands}

\author{Candice~Thomas}
\affiliation{Department of Physics and Astronomy, Purdue University, West Lafayette, Indiana 47907, USA}
\affiliation{Birck Nanotechnology Center, Purdue University, West Lafayette, Indiana 47907, USA}

\author{Sara~Metti}
\affiliation{Birck Nanotechnology Center, Purdue University, West Lafayette, Indiana 47907, USA}
\affiliation{School of Electrical and Computer Engineering, Purdue University, West Lafayette, Indiana 47907, USA}

\author{Charles~R.~Guinn}
\affiliation{Department of Physics and Astronomy, Purdue University, West Lafayette, Indiana 47907, USA}

\author{Ray~Kallaher}
\affiliation{Birck Nanotechnology Center, Purdue University, West Lafayette, Indiana 47907, USA}
\affiliation{Microsoft Quantum at Station Q Purdue, Purdue University, West Lafayette, Indiana 47907, USA}

\author{Mario~Lodari}
\affiliation{QuTech and Kavli Institute of Nanoscience, Delft University of Technology, 2600 GA Delft, The Netherlands}

\author{Giordano~Scappucci}
\affiliation{QuTech and Kavli Institute of Nanoscience, Delft University of Technology, 2600 GA Delft, The Netherlands}

\author{Tiantian~Wang}
\affiliation{Department of Physics and Astronomy, Purdue University, West Lafayette, Indiana 47907, USA}
\affiliation{Birck Nanotechnology Center, Purdue University, West Lafayette, Indiana 47907, USA}

\author{Rosa~E.~Diaz}
\affiliation{Birck Nanotechnology Center, Purdue University, West Lafayette, Indiana 47907, USA}

\author{Geoffrey~C.~Gardner}
\affiliation{Birck Nanotechnology Center, Purdue University, West Lafayette, Indiana 47907, USA}
\affiliation{Microsoft Quantum at Station Q Purdue, Purdue University, West Lafayette, Indiana 47907, USA}

\author{Michael~J.~Manfra}
\affiliation{Department of Physics and Astronomy, Purdue University, West Lafayette, Indiana 47907, USA}
\affiliation{Birck Nanotechnology Center, Purdue University, West Lafayette, Indiana 47907, USA}
\affiliation{School of Electrical and Computer Engineering, Purdue University, West Lafayette, Indiana 47907, USA}
\affiliation{Microsoft Quantum at Station Q Purdue, Purdue University, West Lafayette, Indiana 47907, USA}
\affiliation{School of Materials Engineering, Purdue University, West Lafayette, Indiana 47907, USA}

\author{Srijit~Goswami}\email{S.Goswami@tudelft.nl}
\affiliation{QuTech and Kavli Institute of Nanoscience, Delft University of Technology, 2600 GA Delft, The Netherlands}

\maketitle

\textbf{Two-dimensional electron gases (2DEGs) coupled to superconductors offer the opportunity to explore a variety of quantum phenomena. These include the study of novel Josephson effects\cite{Riwar2016}, superconducting correlations in quantum (spin) Hall systems\cite{Hart2014,Pribiag2015,Wan2015,Amet2016,Lee2017,Finocchiaro_2018}, hybrid superconducting qubits\cite{Casparis2017,Wang2018} and emergent topological states in semiconductors with spin-orbit interaction (SOI)\cite{Rokhinson2012,Nichele2017,Ren2018,Fornieri2018}. InSb is a well-known example of such a strong SOI semiconductor, however hybrid superconducting devices in InSb quantum wells remain unexplored. Here, we interface InSb 2DEGs with a superconductor (NbTiN) to create Josephson junctions (JJs), thus providing the first evidence of induced superconductivity in high quality InSb quantum wells. The JJs support supercurrent transport over several microns and display clear signatures of ballistic superconductivity. Furthermore, we exploit the large Land\'{e} g-factor and gate tunability of the junctions to control the current-phase relation, and drive transitions between the $0$ and $\pi$-states. This control over the free energy landscape allows us to construct a phase diagram identifying these $0$ and $\pi$-regions, in agreement with theory. Our results establish InSb quantum wells as a promising new material platform to study the interplay between superconductivity, SOI and magnetism.}

The combined effect of SOI and a Zeeman field is known to significantly alter the current-phase relation of JJs\cite{Bezuglyi2002,Yokoyama2014,Szombati2016}.
In particular, one expects a complete reversal of the supercurrent (i.e., a $\pi$-JJ)\cite{Hart_2017,Chen2018,Li_2018} when the Zeeman and Thouless energy of the system become comparable.
It was shown recently that such a $\0pi$ transition in a 2D system is in fact accompanied by a topological phase transition\cite{Pientka_2017,Hell_2017,Fornieri2018,Ren2018}. This, combined with the promise of creating scalable topological networks\cite{Nayak_2008,Karzig2016,Plugge2017}, provides a strong motivation to study induced superconductivity in 2DEGs. Key requirements for the semiconductor include low disorder, large SOI and a sizable g-factor, combined with the ability to grow it on the wafer scale. 

\begin{figure*}[!t]
	\includegraphics[scale=1]{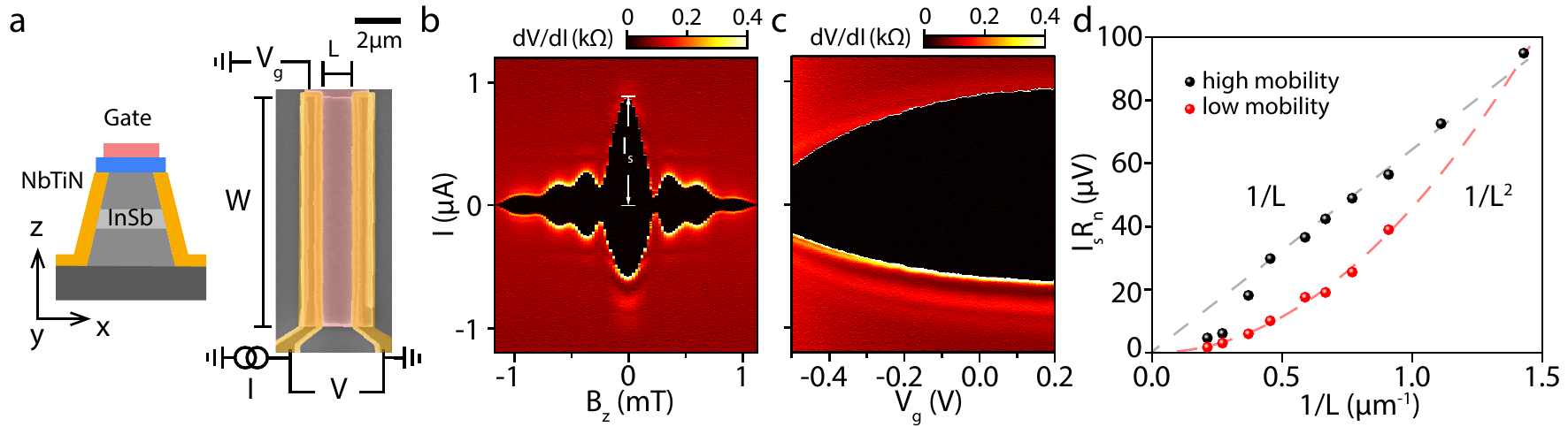} \\
	\caption{\textbf{$|$ Ballistic superconductivity in InSb 2DEGs.} \textbf{a,} Cross-sectional schematic and false-colored scanning electron micrograph (along with a measurement schematic) of a top-gated JJ of width $W$ and length $L$. \textbf{b,} Differential resistance, $\text{d}V/\text{d}I$, versus perpendicular magnetic field, $B_{\text{z}}$, and current bias, $I$, displaying a Fraunhofer-like interference pattern for a JJ with $W=9.7$~$\mu$m, $L=1.1$~$\mu$m. White line indicates the magnitude of the switching current, $I_\text{s}$, at zero magnetic field. \textbf{c,} $\text{d}V/\text{d}I$ as a function of $I$ and gate voltage, $V_\text{g}$, for the same JJ, showing gate control of $I_\text{s}$. \textbf{d,} Length dependence of $I_{\text{s}}R_{\text{n}}$ for JJs on a high mobility (black dots) and low mobility (red dots) wafer. Dashed lines are $1/L$ and $1/L^2$ fits to the data, indicating ballistic and diffusive transport, respectively.}
	\label{fig:1}
\end{figure*}

InSb satisfies all of these requirements\cite{Gilbertson2009,Kallaher2010,Nedniyom2009,Qu_2016} and has emerged as a prime material candidate for engineering topological superconductivity, as evident from nanowire-based systems\cite{Mourik2012,Zhang2017a}. However, despite significant progress in the growth of InSb 2DEGs\cite{Yi2015,Masuda2018}, material challenges have prevented a systematic study of the superconducting proximity effect in these systems. We overcome these issues and reliably create JJs in high mobility InSb quantum wells, allowing us to realize ballistic $\pi$-junctions that can be controlled by magnetic and electric fields.   

The JJs are fabricated in an InSb 2DEG wafer grown by molecular beam epitaxy, with a nominal electron density $n=2.7\cdot10^{11}$~$\text{cm}^{-2}$ and mobility $\mu\approx~150,000~\text{cm}^2/\text{Vs}$, which corresponds to a mean free path $l_{\text{e}}\approx1.3~\mu\text{m}$. Figure~\ref{fig:1}a shows a cross-sectional illustration and scanning electron micrograph of a typical JJ. Following a wet etch of the 2DEG in selected areas, NbTiN is deposited to create side-contacts to the 2DEG, thus defining a JJ of width $W$ and length $L$. Prior to sputtering NbTiN, an in-situ argon plasma cleaning of the exposed quantum well is performed in order to obtain good electrical contacts. A metal top-gate, deposited on a thin dielectric layer is used to modify the electron density in the JJ. Details of the device fabrication and wafer growth can be found in Supplementary Information (SI). 

The junctions are measured using a quasi-four terminal current-biased circuit (Fig.~\ref{fig:1}a) at a temperature of 50~mK. We observe a clear supercurrent branch with zero differential resistance, $\text{d}V/\text{d}I$, followed by a jump to the resistive branch at switching current, $I_{\text{s}}$. In small perpendicular magnetic fields, $B_{\text{z}}$, Fraunhofer-like interference patterns are observed, as seen in Fig.~\ref{fig:1}b. The magnitude of supercurrent is controlled using the gate (Fig.~\ref{fig:1}c). Lowering the gate voltage, $V_{\text{g}}$, leads to a reduction of the electron density in the 2DEG and therefore to a suppression of $I_{\text{s}}$ and an increase in the normal state resistance, $R_{\text{n}}$. In addition, we observe multiple Andreev reflections indicating an induced superconducting gap of 0.9~meV, and excess current measurements allow us to estimate transparencies in the range of 0.6-0.7 (representative data is provided in the SI). 

Studying JJs of varying lengths ($L=0.7-4.7~\mu\text{m}$), we gain insight into the transport regime. These devices fall in the long junction limit, since their lengths exceed the induced superconducting coherence length of around $500~\text{nm}$ (see SI). In this limit the product of the critical current, $I_{\text{c}}$, and $R_{\text{n}}$ is proportional to the Thouless energy\cite{Altshuler1987}, $E_{\text{Th}}=\hbar v_{\text{F}} l_{\text{e}}/2L^2$, where $v_{\text{F}}$ is the Fermi velocity in the 2DEG. Thus, for ballistic (diffusive) transport where $l_{\text{e}}=L$ ($l_{\text{e}}<L$), we expect $I_{\text{c}}R_{\text{n}}$ to scale as $1/L$ ($1/L^2$). Figure~\ref{fig:1}d shows $I_{\text{s}}R_{\text{n}}$\cite{refnoteIs} for a set of JJs. We find a $1/L$ scaling (black dots) indicative of ballistic superconductivity, with deviations only for the longer ($L\geq 2.7~\mu\text{m}$) junctions. Such a $1/L$ dependence up to micron-scale lengths has only recently been experimentally observed in clean graphene-based JJs \cite{Shalom,Borzenets2016}. To confirm the scaling arguments we also include data from a lower mobility wafer with $l_{\text{e}}\approx0.5~\mu\text{m}$ (red dots) and find a $1/L^2$ scaling, consistent with diffusive behavior. In the remainder of this work we focus on JJs fabricated on the high mobility wafer.

\begin{figure*}[!t]
	\includegraphics[scale=1]{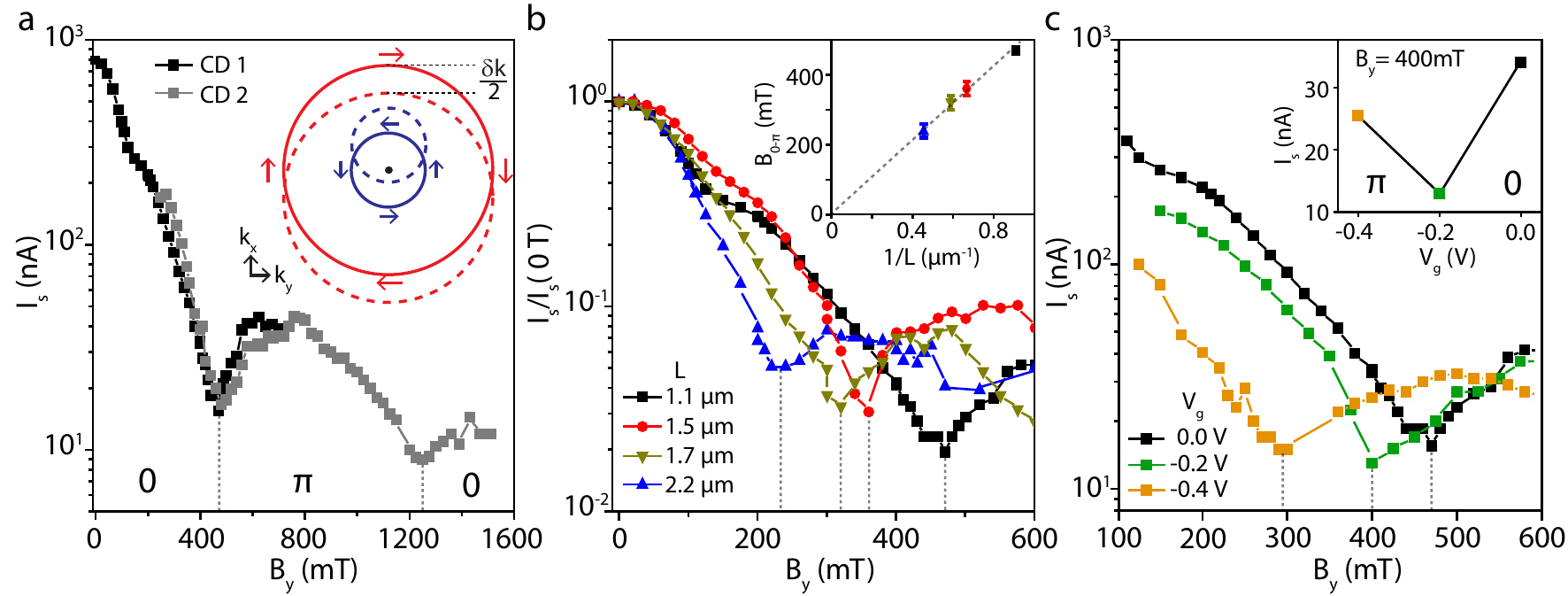}
	\caption{$|$ \textbf{Magnetic field-driven $\0pi$ transitions.} \textbf{a,} Variation of the switching current, $I_{\text{s}}$, with in-plane magnetic field, $B_{\text{y}}$, at $V_\text{g}=0$~V for the same JJ as in Fig.~\ref{fig:1}b,c. Two distinct revivals of $I_{\text{s}}$ are visible at $B_{\text{y}}=470~\text{mT}$ and $1250~\text{mT}$, associated with $\0pi$ transitions. The data is from two cool downs (CDs). The momentum shift, $\delta k/2$, of the Fermi surfaces due to the Zeeman field is sketched in the inset. The solid (dashed) lines depict the situation at zero (finite) magnetic field, and the arrows represent the spin orientation. \textbf{b,} $I_\text{s}$ as a function of $B_{\text{y}}$ at $V_{\text{g}}=0~\text{V}$ for four JJs with different lengths. For better visibility, $I_{\text{s}}$ is normalized with respect to $I_{\text{s}}$ at $B_{\text{y}}=0~\text{T}$. Dashed lines indicate $B_{\0pi}$, the field at which the transition occcurs for each length. The inset shows a linear dependence of $B_{\0pi}$ on $1/L$, in agreement with ballistic transport. \textbf{c,} $I_{\text{s}}$ vs. $B_{\text{y}}$ at three different $V_{\text{g}}$ for the JJ with $L=1.1~\mu\text{m}$. $B_{\0pi}$ shifts to lower values of $B_{\text{y}}$ with more negative gate voltages. $I_{\text{s}}$ vs. $V_{\text{g}}$ at $B_{y}=400~\text{mT}$ shows a non-monotonic behavior as displayed in the inset. The length and gate dependence of panel \textbf{b} and \textbf{c} are in qualitative agreement with Eq.~\ref{eq.1}.}
	\label{fig:2}
\end{figure*}

Using these ballistic junctions, we now explore their response to a Zeeman field. At zero $B$ the Fermi surfaces are split due to the Rashba SOI (solid lines of Fig.~\ref{fig:2}a inset). The magnetic field then splits the bands by the Zeeman energy, $E_\text{Z}\,=\,g \mu_{\text{B}} B$, leading to a shift in the Fermi surfaces~\cite{refnotesoi} by $\pm\delta k/2$. Therefore, the Cooper pairs (electrons with opposite momentum and spin) possess a finite momentum, given by $\bm{k}_{\text{F}} \cdot \delta\bm{k} = E_{\text{Z}}(m^*/\hbar^2)$, where $\bm{k}_{\text{F}}$ is the Fermi momentum and $m^*$ the effective mass.
This translates to a phase acquired by the superconducting order parameter along the direction of current flow, $\Psi(\bm{r})\propto\cos (\delta\bm{k} \cdot \bm{r})$\cite{Bulaevskii_1977,Demler1997,Buzdin_2005}.
Depending on the length of the Cooper pair trajectories, $|\bm{r}|$, the order parameter is either positive or negative, corresponding to the ground state of the JJ being at 0 or $\pi$ superconducting phase difference, respectively.
This oscillation of the order parameter results in a modulation of the critical current $I_{\text{c}} \propto |\Psi|$, where a minimum of $I_{\text{c}}$ is expected whenever the order parameter switches sign\cite{Bezuglyi2002,Yokoyama2014}.
Taking only trajectories perpendicular to the contacts ($\delta\bm{k}\,=\,\delta k \hat{x}$, $\bm{k}_{\text{F}}=k_{\text{F}} \hat{x}$), a JJ with length $L$ will display minima in $I_{\text{c}}$ when $L\delta k\,=\,(2\text{N}+1)\pi/2$, with $N=0,1,2...$ . 
The condition for the first minimum ($N=0$) can be expressed as a resonance condition in terms of the Zeeman and ballistic Thouless energy as $E_{\text{Z}}=\pi E_{\text{Th}}$ giving:
\begin{equation}
g \mu_{\text{B}} B = \pi \frac{\hbar^2 \sqrt{2 \pi n}}{m^* 2L}.
\label{eq.1}
\end{equation}
The $\0pi$ transition therefore depends on three experimentally accessible parameters: 1) applied magnetic field, 2) length of the JJ and 3) carrier density. In the following, we demonstrate independent control of each of these parameters, allowing for a complete study of the free energy landscape of the junctions. 

\begin{figure}[!t]
	\includegraphics[scale=1]{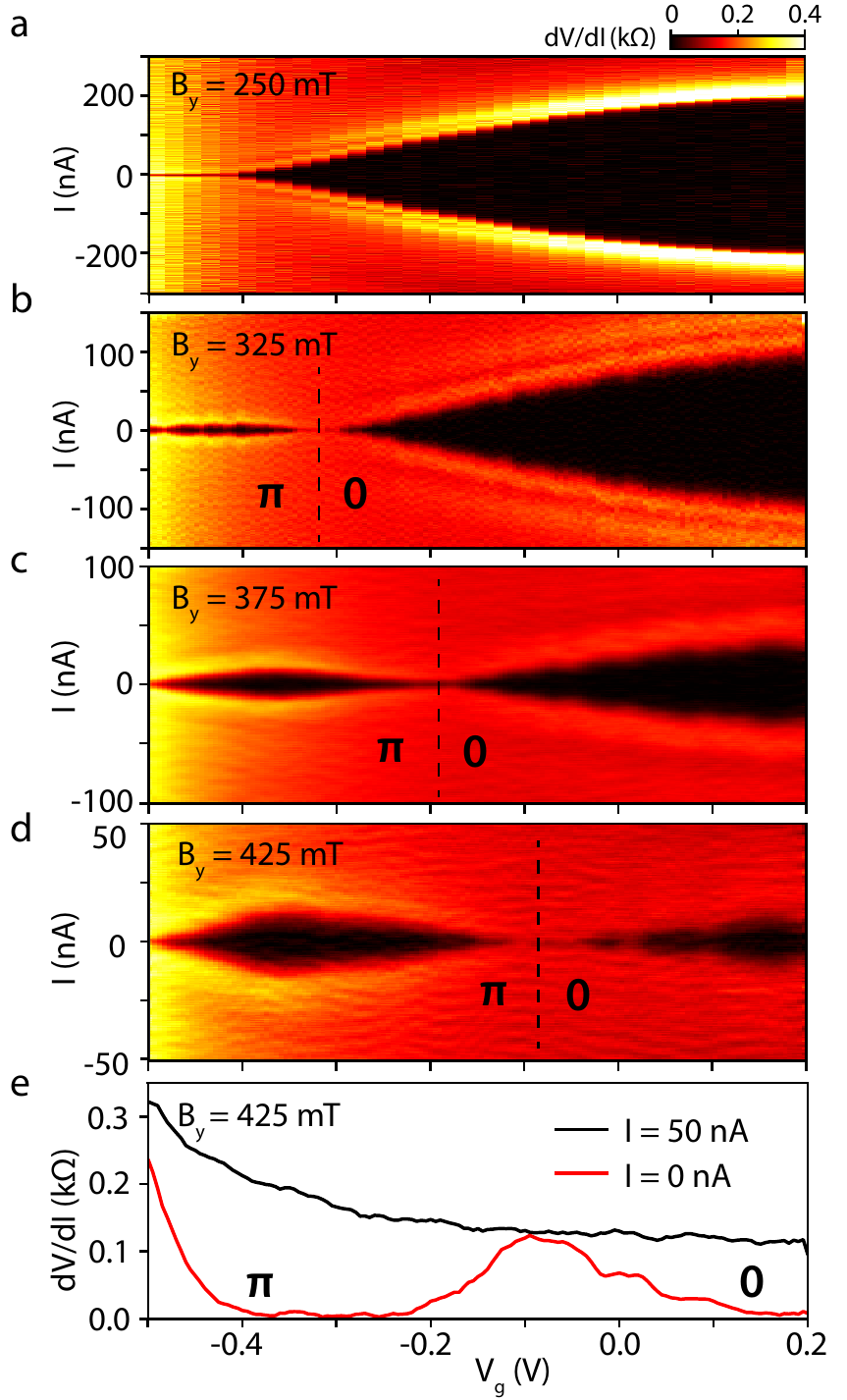}
	\caption{\textbf{$|$ Gate-driven $\0pi$ transitions.} \textbf{a-d,} $\text{d}V/\text{d}I$ as a function of $I$ and $V_{\text{g}}$ for several $B_y$ as indicated. From $B_{y}=325~\text{mT}$ onward, a gate-driven $\0pi$ transition becomes evident, characterized by a re-emergence of $I_{\text{s}}$ with decreasing $V_{\text{g}}$. As expected, the transition shifts to higher gate voltages with increasing $B_{\text{y}}$ (see SI for sweeps at additional values of the magnetic field). \textbf{e,} Line-cuts through panel \textbf{d} at $I=50~\text{nA}$ (black) and $I=0~\text{nA}$ (red). The low bias trace reveals the $\0pi$ transition whereas the high bias trace shows a monotonic behavior.}
	\label{fig:3}
\end{figure}

We start by varying $B$, while $n$ (controllable by $V_{\text{g}}$) and $L$ remain fixed.
Figure~\ref{fig:2}a shows the expected oscillation of $I_{\text{s}}$ with increasing $B_{\text{y}}$, displaying two distinct minima at $B_{\text{y}}=470~\text{mT}$ and $B_{\text{y}}=1250~\text{mT}$ (see SI for details about magnetic field alignment). 
This behavior is consistent with a magnetic field driven $\0pi$ transition, as discussed above, where the first (second) minimum corresponds to a transition of the JJ state from $0$ to $\pi$ ($\pi$ to $0$). This interpretation is corroborated by the occurrence of the second minimum at a field value which is approximately three times larger than the first. 
Note that this is incompatible with a Fraunhofer interference pattern that might arise from the finite thickness of the 2DEG. Furthermore, taking into account the gate dependence of the transition and other geometric considerations (discussed in detail in the SI) allows us to conclusively rule out such a mechanism for the supercurrent modulation.

Next, we investigate how the length of the JJ influences $B_{\0pi}$, the magnetic field at which the transition occurs. Figure~\ref{fig:2}b presents the $I_{\text{s}}$ oscillation for JJs with four different lengths, showing that $B_{\0pi}$ is systematically reduced for increasing $L$. 
Plotting $B_{\0pi}$ with respect to $1/L$ (inset of Fig.~\ref{fig:2}b), we find a linear dependence as expected from Eq.~\ref{eq.1}.
The transition points are therefore determined by the ballistic $E_{\text{Th}}$, consistent with the conclusions from Fig.~\ref{fig:1}d. Finally, we check the dependence of the transition on the electron density. 
In Fig.~\ref{fig:2}c, we plot $I_{\text{s}}$ versus $B_{\text{y}}$ for different gate voltages using a JJ with $L=1.1~\mu\text{m}$.
As $V_{\text{g}}$ is lowered, $B_{\0pi}$ shifts to smaller values, again in qualitative agreement with Eq.~\ref{eq.1}. Interestingly, above a certain magnetic field the state of the JJ ($0$ or $\pi$) becomes gate-dependent. For example at $B_{\text{y}}=400~\text{mT}$, the junction changes from a 0-JJ ($V_{\text{g}}=0~\text{V}$) to a $\pi$-JJ ($V_{\text{g}}=-0.4~\text{V}$), with a transition at $V_{\text{g}}=-0.2~\text{V}$. This indicates the feasibility of tuning the JJ into the $\pi$-state using gate voltages, while the magnetic field remains fixed.

These gate-driven transitions are demonstrated in Fig.~\ref{fig:3}a-d, which show a sequence of $I-V_{\text{g}}$ plots for increasing in-plane magnetic fields. At $B_{\text{y}}=250~\text{mT}$, $I_{\text{s}}$ displays a monotonic reduction with decreasing $V_{\text{g}}$. At a higher magnetic field, $B_{\text{y}}=325$~mT, $I_{\text{s}}$ reveals a markedly different behavior, whereby the supercurrent first decreases and then (at $V_{\text{g}}=-0.32~\text{V}$) shows a clear revival, indicative of a gate-driven $\0pi$ transition, where the resonance condition ($E_{\text{Z}}=\pi E_{\text{Th}}$) is achieved by tuning the electron density. Increasing $B_{\text{y}}$ further, continuously moves the transition point to higher gate voltages (larger density), perfectly in line with expectations for a $\0pi$ transition. Figure~\ref{fig:3}e shows two line-cuts from Fig.~\ref{fig:3}d. At zero current bias, $\text{d}V/\text{d}I$ shows a clear peak, indicative of a re-entrance of the supercurrent due to the the $\0pi$ transition. However, at high bias, $\text{d}V/\text{d}I$ increases monotonically, similar to the response at zero magnetic field. This eliminates trivial interference effects as an explanation for the supercurrent modulation, where one would expect a correlation between the two curves~\cite{Calado2015,Shalom,Allen2017}. 

\begin{figure}[!t]
	\includegraphics[scale=1]{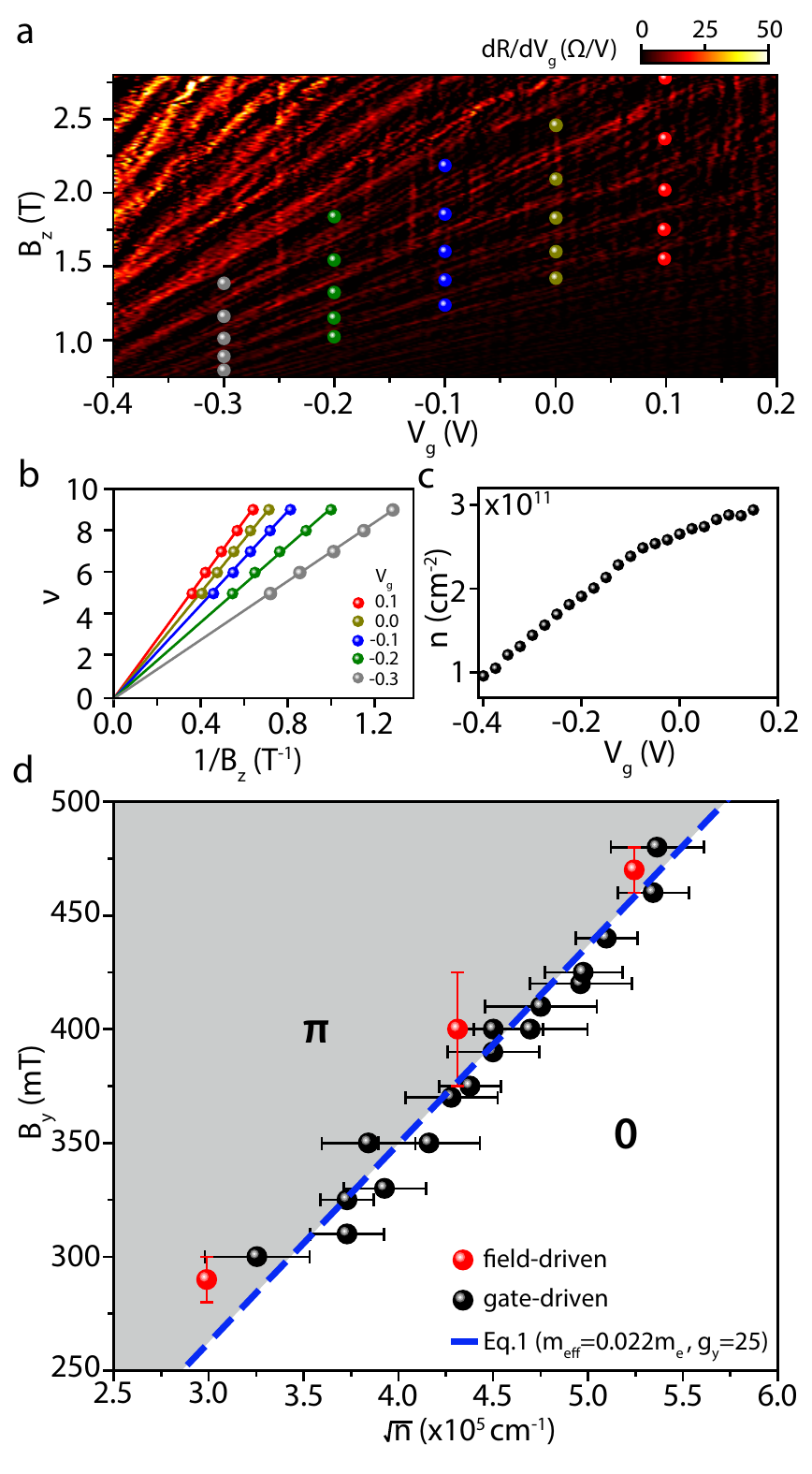}
	\caption{\textbf{$|$ $\0pi$ phase diagram.} \textbf{a,} Landau fan diagram for the JJ with $L=1.1~\mu\text{m}$, showing the transresistance ($\text{d}R/\text{d}V_{\text{g}}$) as a function of $B_{\text{z}}$ and $V_{\text{g}}$. The symbols indicate positions of integer filling factors $\nu$ at specific values of $V_{\text{g}}$. \textbf{b,} Dependence of $\nu$ on $1/B_{\text{z}}$ along with linear fits used to extract the electron density, $n (V_\text{g})$, presented in \textbf{c}. \textbf{d,} Phase diagram of the $\0pi$ transition as a function of $B_{\text{y}}\propto E_{\text{Z}}$ and $\sqrt{n}\propto E_{\text{Th}}$, containing all data points obtained from both field-driven (red) and gate-driven (black) $\0pi$ transitions. For the error analysis, see SI. We fit the data  to Eq.~\ref{eq.1} (blue line) with $g_y$ as a fitting parameter.}
	\label{fig:4}
\end{figure}

In contrast to the field-driven measurements (Fig.~\ref{fig:2}), controlling the transition with a gate avoids the need for time-consuming field alignment procedures, thus allowing us to efficiently explore a large parameter space in magnetic field and gate voltage. We now combine these results to construct a $\0pi$ phase diagram of the JJ. The combination of a high quality 2DEG and relatively long devices results in well defined magneto-resistance oscillations, allowing us to directly extract the electron density in the junction. Figure~\ref{fig:4}a shows the Landau fan diagram in perpendicular magnetic fields, $B_{\text{z}}$, from which we identify the filling factors, $\nu = nh/eB_{\text{z}}$ (Fig.~\ref{fig:4}b), and thereby obtain the $n$ vs. $V_{\text{g}}$ curve (Fig.~\ref{fig:4}c). We then plot all the transition points in Fig.~\ref{fig:4}d. The axes represent the two important energy scales in the system ($B_{\text{y}}\propto E_{\text{Z}}$ and $\sqrt{n}\propto E_{\text{Th}}$), thereby highlighting the $0$ and $\pi$ regions in the phase space. Finally, we compare our results with the theory of ballistic JJs represented by Eq.~1. To do so, we independently extract the effective mass (see SI), $m^*= (0.022\pm 0.002) m_{\text{e}}$, and fit the data to a single free parameter, $g_y$ (the in-plane g-factor), giving  $g_y=25\pm 3$ in good agreement with previous measurements on similar InSb quantum wells~\cite{Qu_2016}.

Our work provides the first evidence of induced superconductivity in high quality InSb 2DEGs and demonstrates the creation of robust, gate-tunable $\pi$-Josephson junctions. We show that the $\0pi$ transition can be driven both by magnetic fields and gate voltages. The significant region of phase space where the $\pi$-JJ is stable could prove advantageous in the study of topological superconductivity in planar JJs~\cite{Pientka_2017,Hell_2017,Fornieri2018,Ren2018}. Moreover, these large SOI 2DEGs, in conjunction with our magnetic field compatible superconducting electrodes and clear Landau quantization, would also be excellent candidates to realize topological junctions in the quantum Hall regime\cite{Finocchiaro_2018}. Finally, the ability to control the ground state between $0$ and $\pi$ states using gates is analogous to recent experimental results in ferromagnetic JJs\cite{Gingrich_2016}, and could possibly serve as a semiconductor-based platform for novel superconducting logic applications~\cite{Terzioglu1998}. We therefore establish InSb 2DEGs as a new, scalable platform for developing hybrid superconductor-semiconductor technologies.         
\\

\newpage
\textbf{Acknowledgments:} We thank Ady Stern, Attila Geresdi and Michiel de Moor for useful discussions. The research at Delft was supported by the Dutch National Science Foundation (NWO) and a TKI grant of the Dutch topsectoren program. The work at Purdue was funded by Microsoft Quantum. 
\\
\\
\textbf{Author contributions:} C.~T.~K. and C.~M.~M. fabricated and measured the devices. C.~T., G.~C.~G. and M.~J.~M. designed and grew the semiconductor heterostructures. C.~T., S.~M., C.~R.~G., R.~K., T.~W., R.~E.~D., G.~C.~G., and M.~J.~M. characterized the materials. M.~L. and G.~S. provided the effective mass measurements. C.~T.~K., C.~M.~M., F.~K.~d.~V. and S.~G. performed the data analysis. The manuscript was written by C.~T.~K., F.~K.~d.~V., C.~M.~M., and S.~G., with input from all co-authors. S.~G. supervised the project.

\clearpage
\newcommand{\beginsupplement}{%
	\setcounter{section}{0}
	\renewcommand{\thesection}{S\arabic{section}}%
	\setcounter{table}{0}
	\renewcommand{\thetable}{S\arabic{table}}%
	\setcounter{figure}{0}
	\renewcommand{\thefigure}{S\arabic{figure}}%
	\setcounter{equation}{0}
	\renewcommand{\theequation}{S\arabic{equation}}%
}
\onecolumngrid

\section*{\large{Supplementary Information}}
\vspace{10mm}

\beginsupplement

\section*{Wafer growth and characterization}
\begin{figure}[!ht]
	\includegraphics[width=.8\textwidth]{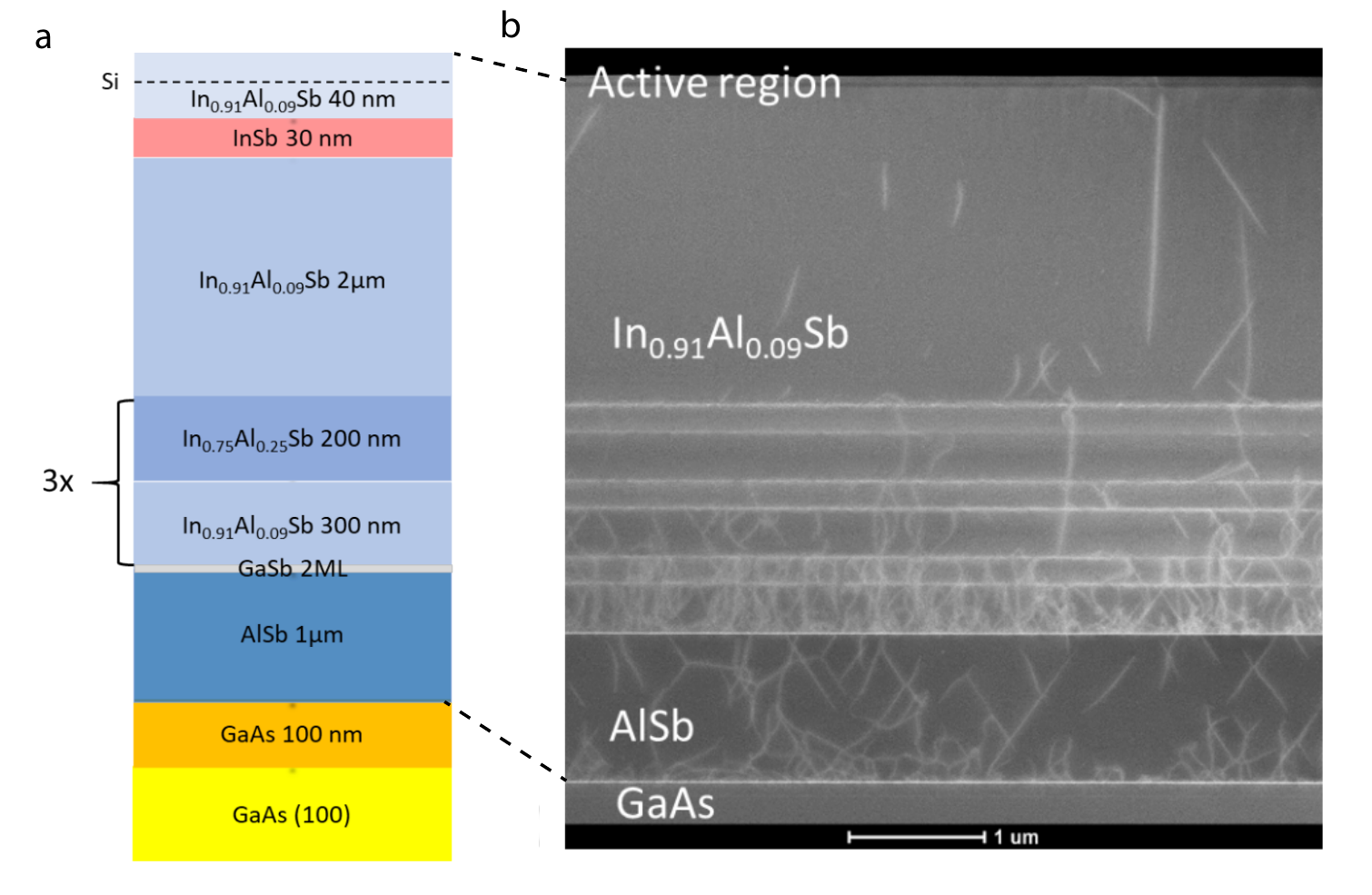}
	\caption{(a), Layer stack of the InSb/GaAs heterostructure, where the layer constituents and thicknesses are indicated. (b), Scanning transmission electron micrograph of the structure in (a), obtained in High Angle Annular Dark Field Mode along the [110] zone axis.}
	\label{SF.1}
\end{figure} 
InSb-based 2DEGs were grown on semi-insulating GaAs (100) substrates by molecular beam epitaxy in a Veeco Gen 930 using ultra-high purity techniques and methods as described in Ref.~\citenum{Gardner2016}. The layer stack of the heterostructure is shown in Fig.~\ref{SF.1}a. The growth has been initiated with a $100\,\text{nm}$ thick GaAs buffer followed by a $1\,\mu\text{m}$ thick AlSb nucleation layer. The metamorphic buffer is composed of a superlattice of $300\,\text{nm}$ thick In$_{0.91}$Al$_{0.09}$Sb and $200\,\text{nm}$ thick In$_{0.75}$Al$_{0.25}$Sb layers, repeated 3 times, and directly followed by a $2\,\mu\text{m}$ thick In$_{0.91}$Al$_{0.09}$Sb layer. The active region consists of a $30\,\text{nm}$ thick InSb quantum well and a $40\,\text{nm}$ thick In$_{0.91}$Al$_{0.09}$Sb top barrier. The Si $\delta$-doping layer has been introduced at $20\,\text{nm}$ from the quantum well and the surface. The In$_\text{x}$Al$_{1-\text{x}}$Sb buffer, the InSb quantum well and the In$_\text{x}$Al$_{1-\text{x}}$Sb setback were grown at a temperature of $440$~\textcelsius\ under a p(1x3) surface reconstruction. The growth temperature was lowered to $340$~\textcelsius, where the surface reconstruction changed to c(4x4), just before the $\delta$-doping layer, to facilitate Si incorporation \cite{Liu1998}. 
The scanning transmission electron micrograph of Fig.~\ref{SF.1}b reveals the efficiency of the metamorphic buffer to filter the dislocations.\\
The wafer is characterized by measuring the (quantum) Hall effect in a Hall bar geometry at $T=300~\text{mK}$. From a linear fit to the transversal resistance in a magnetic field range up to 1~T, we extract an electron density $n=2.71\cdot 10^{11}~\text{cm}^{-2}$, and by using the longitudinal resistivity at zero field, we obtain a mobility $\mu=146,400~\text{cm}^2/\text{Vs}$ (see Tab. \ref{tab:1}). We calculate the corresponding mean free path to be $l_{\text{e}}=1.26~\mu\text{m}$. In Tab. \ref{tab:1}, we also include $n$, $\mu$ and $l_e$ for the low mobility wafer, obtained from a quantum Hall measurement on this wafer. Data from the low mobility wafer is shown in Fig. 1d in the main text. 

\begin{table}[H]
	\begin{center}
		\begin{tabular}{|c|c|c|} 
			\hline
			&High mobility wafer&Low mobility wafer\\
			\hline\hline
			$n~(\text{cm}^{-2})$&$2.71\cdot10^{11}$&$2.71\cdot10^{11}$\\
			\hline
			$\mu~(\text{cm}^{2}/\text{Vs})$&146,400&61,500\\
			\hline
			$l_{\text{e}}~(\mu\text{m})$&1.26&0.53\\
			\hline
		\end{tabular}
	\end{center}
	\caption{Electron density, mobility and mean free path for the high and low mobility wafer, obtained from quantum Hall measurements at $T=300~\text{mK}$.}
	\label{tab:1}
\end{table}

\section*{Device fabrication}
The devices are fabricated using electron beam lithography. First, mesa structures of width $W$ and length $L$ are defined by etching the InSb 2DEG in selected areas. We use a wet etch solution consisting of 560~ml deionized water, 9.6~g citric acid powder, 5~ml $\text{H}_{2}\text{O}_{2}$ and 7~ml $\text{H}_3\text{PO}_4$, and etch for 5~min, which results in an etch depth around $150~\text{nm}$. This is followed by the deposition of superconducting contacts in an ATC 1800-V sputtering system. Before the deposition, we clean the InSb interfaces in an Ar plasma for 3 min (using a power of 100~W and a pressure of 5~mTorr). Subsequently, without breaking the vacuum, we sputter NbTi (30~s) and NbTiN (330~s) at a pressure of 2.5~mTorr, resulting in a layer thickness of approximately $200~\text{nm}$. Next, a $45~\text{nm}$ thick layer of $\text{AlO}_{\text{x}}$ dielectric is added by atomic layer deposition at 105~\textcelsius, followed by a top-gate consisting of 10~nm/170~nm of Ti/Au.

\section*{Multiple Andreev reflections and excess current}
\begin{figure*}[h]
	\includegraphics[scale=1.0]{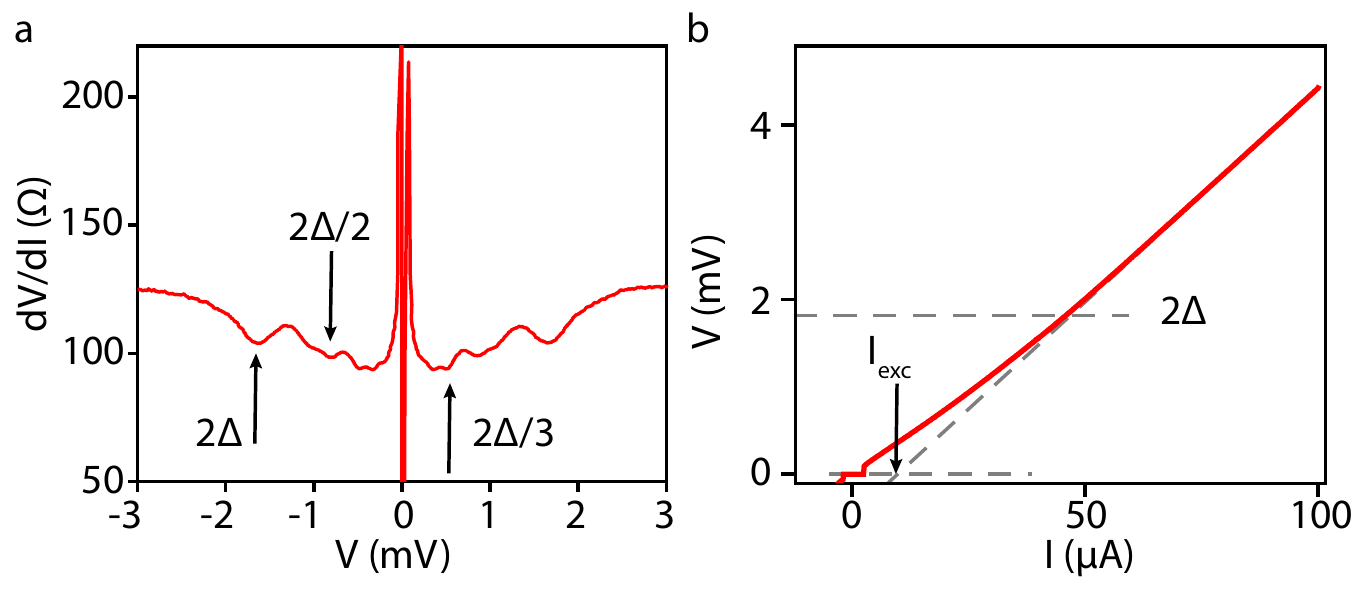}
	\caption{(a), Differential resistance, $\text{d}V/\text{d}I$, as a function of bias voltage, $V$, showing multiple Andreev reflections. Three dips at $V=2\Delta,\, 2\Delta/2 \text{ and } 2\Delta/3$ are highlighted. (b), Voltage measured as a function of bias current. The excess current, $I_\text{exc}$, and $V=2\Delta$ are indicated.}
	\label{SF.2}
\end{figure*}
To further characterize the superconductivity in our JJs, we study multiple Andreev reflections (MAR) in a representative JJ, by measuring its differential resistance, $\text{d}V/\text{d}I$, as a function of applied bias voltage, $V$. 
In Fig.~\ref{SF.2}a, we observe three dips in $\text{d}V/\text{d}I$, the first, at $2\Delta$, corresponding to the coherence peaks of the superconducting density of states, and two MAR peaks at $2\Delta/2$ and $2\Delta/3$. From these peaks we extract an induced superconducting gap $\Delta=0.9~\text{meV}$. 
In addition, we estimate the transparancy of the same JJ by measuring its excess current, $I_\text{exc}$, and normal state resistance, $R_\text{n}$.
This measurement is shown in Fig.~\ref{SF.2}b, where we perform a linear fit in the high bias region of the $I-V$ curve ($V>2\Delta$) and obtain $I_\text{exc}=9~\mu\text{A}$ and $R_\text{n}=50~\Omega$.
Using the OBTK model \cite{Flensberg1988}, we find a value of 0.62 for the transparency of the JJ.

\section*{Weak anti-localization and spin-orbit interaction energy}

\begin{figure*}[!h]
	\includegraphics[width=.5\textwidth]{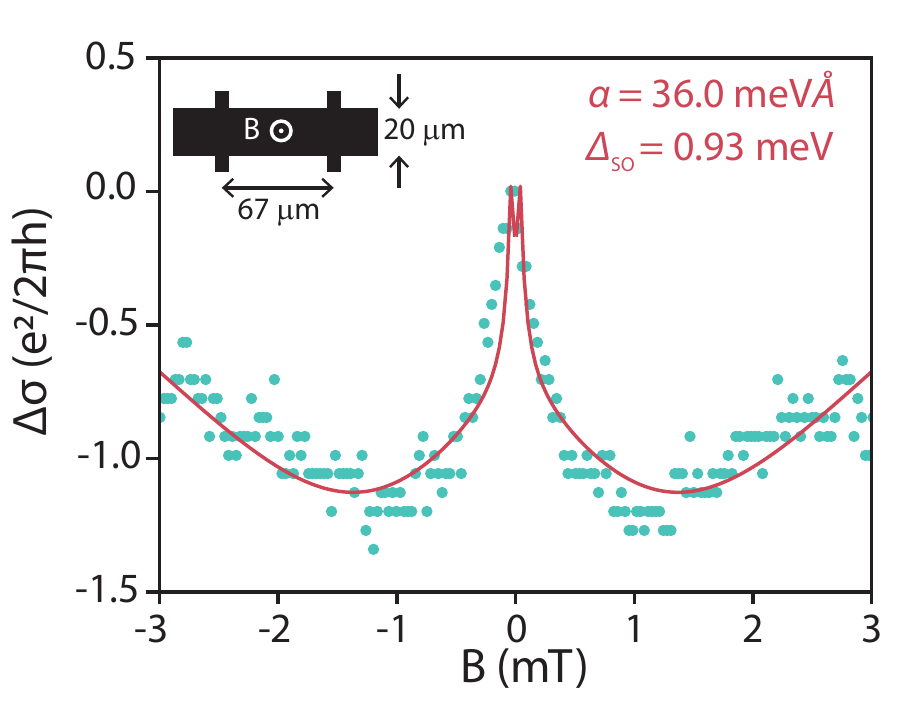} 
	\caption{Measured longitudinal conductivity difference, $\Delta\sigma$, as a function of magnetic field, $B$, displaying a weak anti-localization peak around zero field. We fit (red) the data (cyan) using the ILP model and extract the SOI energy at the Fermi energy, $\Delta_\text{SO}$, from which we calculate the Rashba spin-orbit parameter $\alpha$. The inset shows a schematic of the Hall bar device, indicating its length and width, and the magnetic field direction.}
	\label{SF.3}
\end{figure*}
To obtain an estimate of the typical energy scale associated with the spin-orbit interaction, we performed weak anti-localization (WAL) measurements.
We use a Hall bar device (inset Fig.~\ref{SF.3}) fabricated on the high mobility wafer (Fig.~\ref{SF.1}), and apply magnetic field perpendicular to the Hall bar.
The measurement in Fig.~\ref{SF.3} reveals the typical WAL peak around zero field. 
This peak is caused by suppression of coherent backscattering at small magnetic fields due to the spin-orbit interaction. 
As we expect the Dyakonov Perel scattering mechanism to be dominating in our high mobility wafer, we use the theory developed by Iordanskii, Lyanda-Geller and Pikus~\cite{Iordanskii1994} to fit the data:
\begin{equation*}
\begin{aligned}
\dfrac{\Delta\sigma(B)}{e^2/2\pi h}\,=&\,-\dfrac{1}{a}-\dfrac{2a_0+1+H_\text{s}}{a_1(a_0+H_\text{s}-2H_\text{s})} - 2\ln H_\text{tr} - \Psi\left( 1/2+H_{\phi} \right) - 3C \\
& + \sum_{n=1}^{\infty}\left[\dfrac{3}{n}-\dfrac{3a^2_\text{n}+2a_\text{n}H_\text{s}-1-2(2n+1)H_\text{s}}{(a_\text{n}+H_\text{s}) a_\text{n-1} a_\text{n+1} - 2 H_\text{s} [(2n+1) a_\text{n}-1]}\right] \text{,}
\end{aligned}
\end{equation*}
where $\Psi$ is the Digamma function, $C$ the Euler constant, and 
\begin{equation*}
a_\text{n}=n+\dfrac{1}{2}+H_{\phi}+H_\text{s} \quad \quad H_{\text{tr,$\phi$,s}}=\dfrac{\hbar}{4 e D B \tau_{\text{tr,$\phi$,s}}} \quad \quad\Delta_\text{SO}=\sqrt{\dfrac{2\hbar^2}{\tau_\text{tr} \tau_\text{s}}},
\end{equation*}
with $D=v_\text{F} l_e /2$, and $\tau_{\text{tr,$\phi$,s}}$ the scattering times for elastic, inelastic and spin-orbit scattering, respectively.
We find a spin-orbit energy splitting at the Fermi level ($\Delta_\text{SO}$) of 0.93~meV. 
The Rashba spin-orbit parameter of $\alpha=36~\text{meV}\text{\AA}$ is calculated following $\alpha=\Delta_{\text{SO}}/k_\text{F}$, where $k_F$ is deduced from a classical Hall measurement.
Finally, we compare $\Delta_\text{SO}$ to the Zeeman energy. 
For a Land\'{e} g-factor of 25, $\Delta_{\text{SO}}\,>\, E_{\text{Z}}$ up to 640\,mT. We are therefore in the spin-orbit dominated regime for the $\0pi$ transition.

\section*{Magnetic field alignment}
\begin{figure*}[h]
	\includegraphics[width=\textwidth]{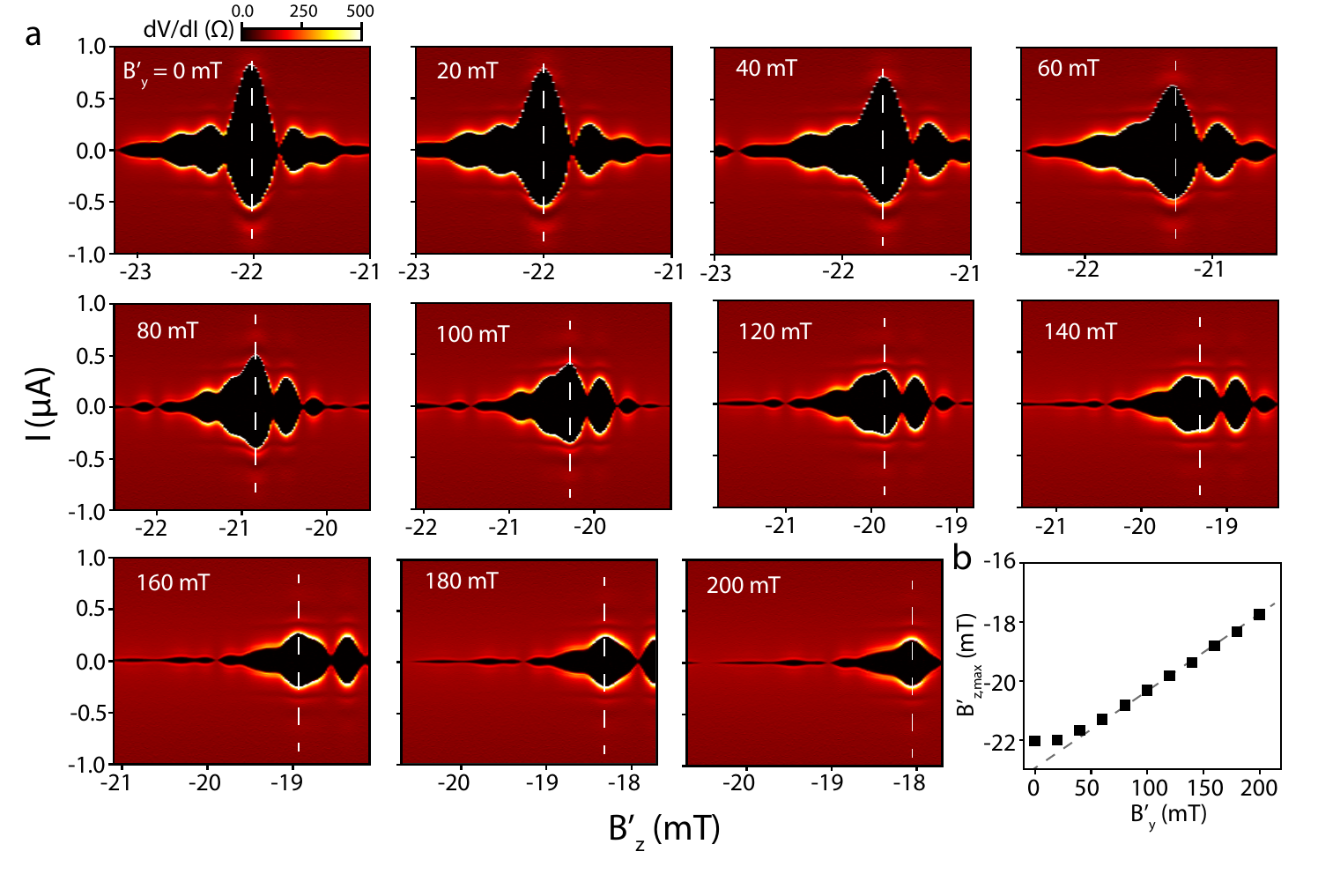}
	\caption{(a), Differential resistance, $\text{d}V/\text{d}I$ maps as a function of current bias, $I$, and out-of-plane magnetic field, $B'_\text{z}$, with increasing in-plane magnetic field, $B'_{\text{y}}$, in steps of $20$~mT. We track the central lobe of the interference pattern, labeled by white dash lines, to obtain $B_{\text{z,max}}$. (b), The $B'_{\text{z,max}}$ vs. $B'_{\text{y}}$ dependence showing the small perpendicular component of $B'_{\text{y}}$.}
	\label{SF.4}
\end{figure*}
To ensure we are sweeping the magnetic field in the plane of the JJs only, we characterize the misalignment of our vector magnet axes, $B'_\text{y}$ and $B'_\text{z}$, used to apply the magnetic field in-plane and out-of-plane of the JJ, $B_\text{y}$ and $B_\text{z}$. 
In Fig.~\ref{SF.4}a we present a systematic measurement of the Fraunhofer interference pattern induced by $B'_\text{z}$ with increasing $B'_{\text{y}}$.
We track the magnetic field at which the central lobe reaches its maximum $I_\text{s}$, $B'_\text{z,max}$ and plot this for all $B'_\text{y}$ in Fig.~\ref{SF.4}b.
The linear dependence observed, represents a small misalignment angle of $\theta = 1.4^\circ$.
We take this angle into account when sweeping the in-plane field, $B_\text{y}=\cos(\theta) B'_\text{y}+\sin(\theta) B'_\text{z}$, and disregard it for the out-of-plane direction, $B_\text{z}=B'_\text{z}$.

\section*{In-plane interference considerations}
\begin{figure*}[!b]
	\includegraphics[width=1\textwidth]{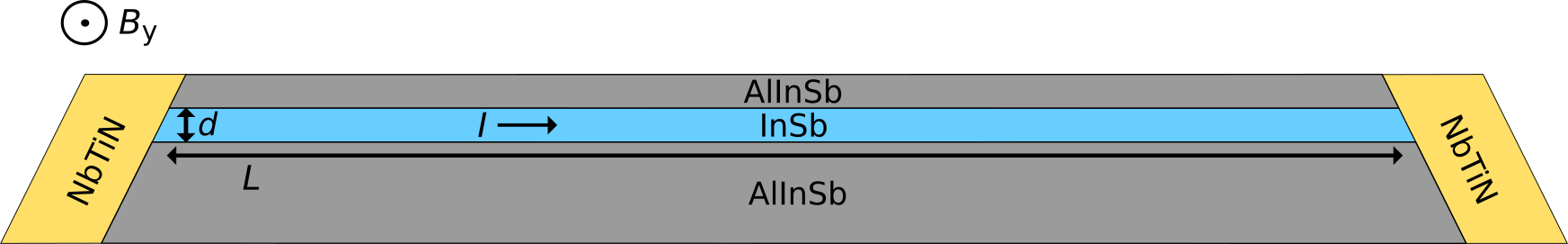}
	\caption{Cross-sectional illustration of the InSb quantum well for a JJ with $L=1.1~\mu\text{m}$ and $d=30~\text{nm}$. The image is drawn to scale and the in-plane magnetic field direction, $B_\text{y}$ is indicated.}
	\label{SF.FH}
\end{figure*}
We observe a switching current, $I_\text{s}$ modulation in a JJ with $L=1.1$~$\mu$m, with minima at 470~mT and 1250~mT, which are attributed to Zeeman induced $\0pi$ transitions.
One might be inclined to believe that this modulation is caused by an in-plane Fraunhofer interference effect, due to the finite thickness ($d=30\,\text{nm}$) of the InSb quantum well. 
The $I_s$ minima of such a Fraunhofer pattern are expected to occur at $B_{\text{node}}=\text{N}\Phi_0/A$, where $\Phi_0$ is the magnetic flux quantum, $A=d\cdot L$ is the cross-sectional 2DEG area and $\text{N}=1,2,3,...$. 
The second minimum should thus occur at twice the value of the first, which is not the case here. 
Moreover, based on the estimated cross-sectional area of the JJ (see Fig.~\ref{SF.FH}), one would expect the first node to be at 60~mT, inconsistent with the observation.
In fact, it has been shown~\cite{Monaco2009,Chiodi2012} that a oscillatory interference pattern is not expected at all in such an SNS junction with $L\gg d$.
Finally, for an in-plane interference effect one expects the $B$ value at which the $I_s$ minima occur to increase for more negative gate voltages, since the wavefunction is then squeezed and $d$ effectively reduced. However, we observe the opposite behavior (i.e., the minima move to lower $B$), as expected for Zeeman-induced $\0pi$ transitions. To conclude, we rule out an in-plane interference effect as a possible explanation for the supercurrent modulation. 

\section*{Additional gate-driven $\0pi$ transitions}
\begin{figure*}[h]
	\includegraphics[width=.98\textwidth]{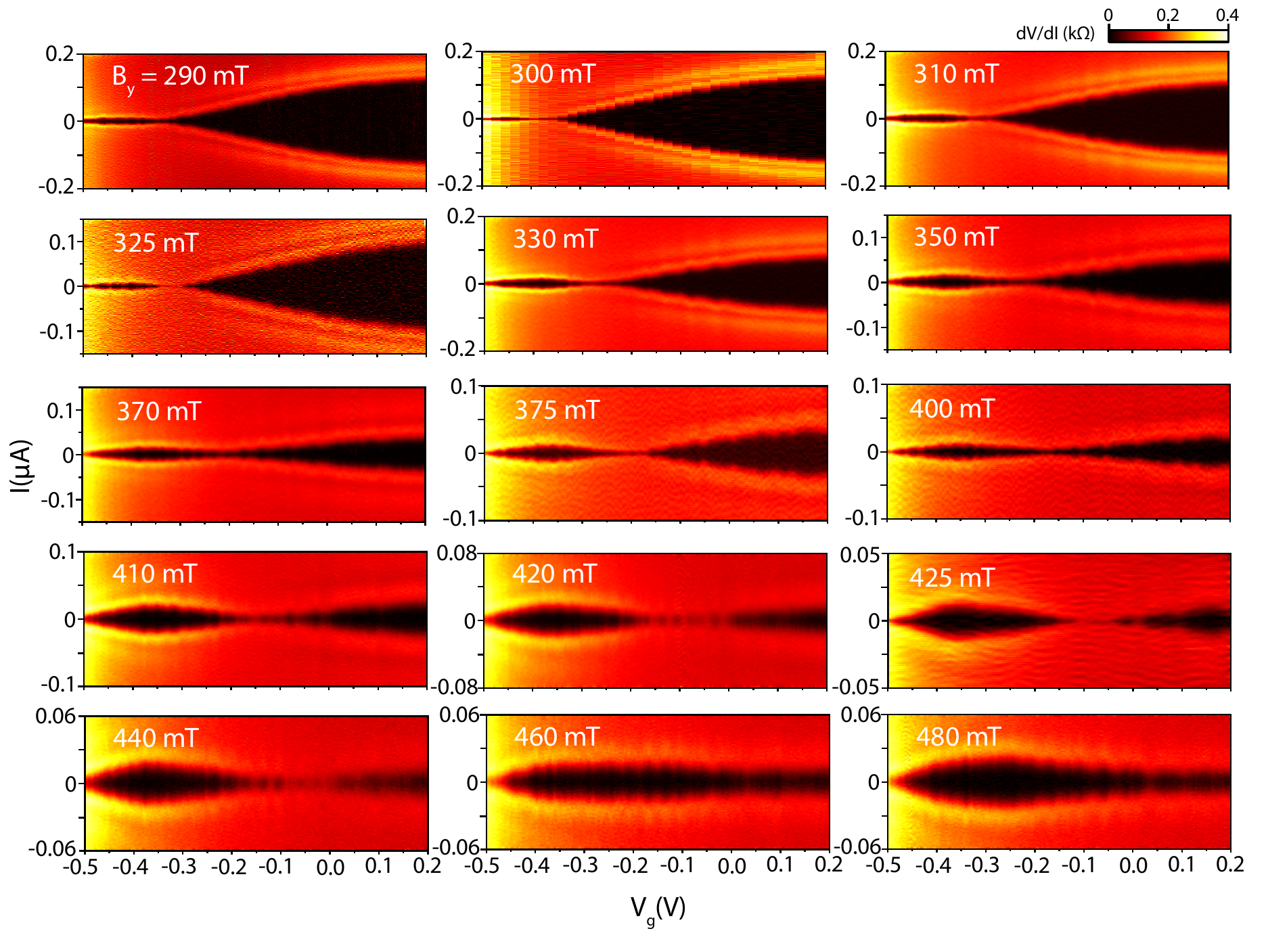}
	\caption{Differential resistance, $\text{d}V/\text{d}I$, as a function of current bias, $I$, and gate voltage, $V_\text{g}$, for the in-plane magnetic field values, $B_\text{y}$, indicated.}
	\label{SF.5-1}
\end{figure*}
Here, we present additional data of the gate-driven $\0pi$ transitions in the JJ with $L=1.1~\mu\text{m}$. The gate voltages of the $\0pi$ transitions presented in the phase diagram are extracted from the plots in Fig.~\ref{SF.5-1}.

\section*{Error analysis for gate-driven $\0pi$ transitions}
\begin{figure*}[!t]
	\includegraphics[scale=0.8]{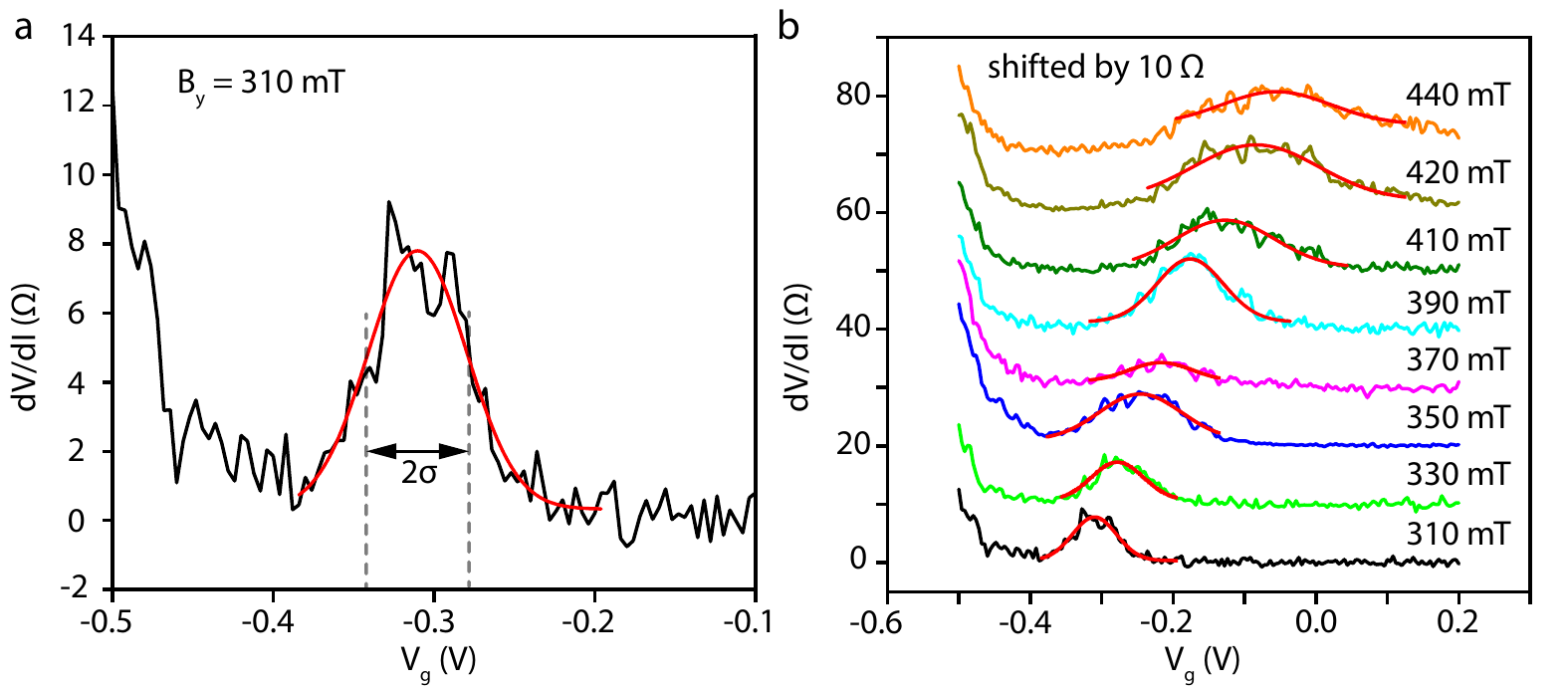}
	\caption{(a-b), Linetraces of the differential resistance, $\text{d}V/\text{d}I$, in the JJ with $L=1.1~\mu\text{m}$ as a function of gate voltage, $V_\text{g}$, for magnetic fields, $B_\text{y}$ of $310$~mT for (a), and as indicated for (b), respectively. The peaks observed are fitted with a Gaussian curve, to obtain the standard deviation, $\sigma$. In (b) the traces are shifted for clarity.}
	\label{SF.5}
\end{figure*}
To systematically extract the value where gate-driven $\0pi$ transition occurs and its error, we use a fit of the linetraces from Fig.~\ref{SF.5-1}, at zero $I$. 
At the transition point, a peak in $\text{d}V/\text{d}I$ indicates the $\0pi$ transition.
As an example, we show a single linetrace at $310$~mT in Fig.~\ref{SF.5}a, and extract the standard deviation, $\sigma$, based on a Gaussian fit of the peak. Subsequently, we used the gate to density mapping to convert $\sigma$ to the error bar shown in the phase diagram. This fitting procedure is used for all magnetic fields (Fig.~\ref{SF.5}b). 

\section*{Effective mass measurement}

\begin{figure*}[!b]
	\includegraphics[scale=0.8]{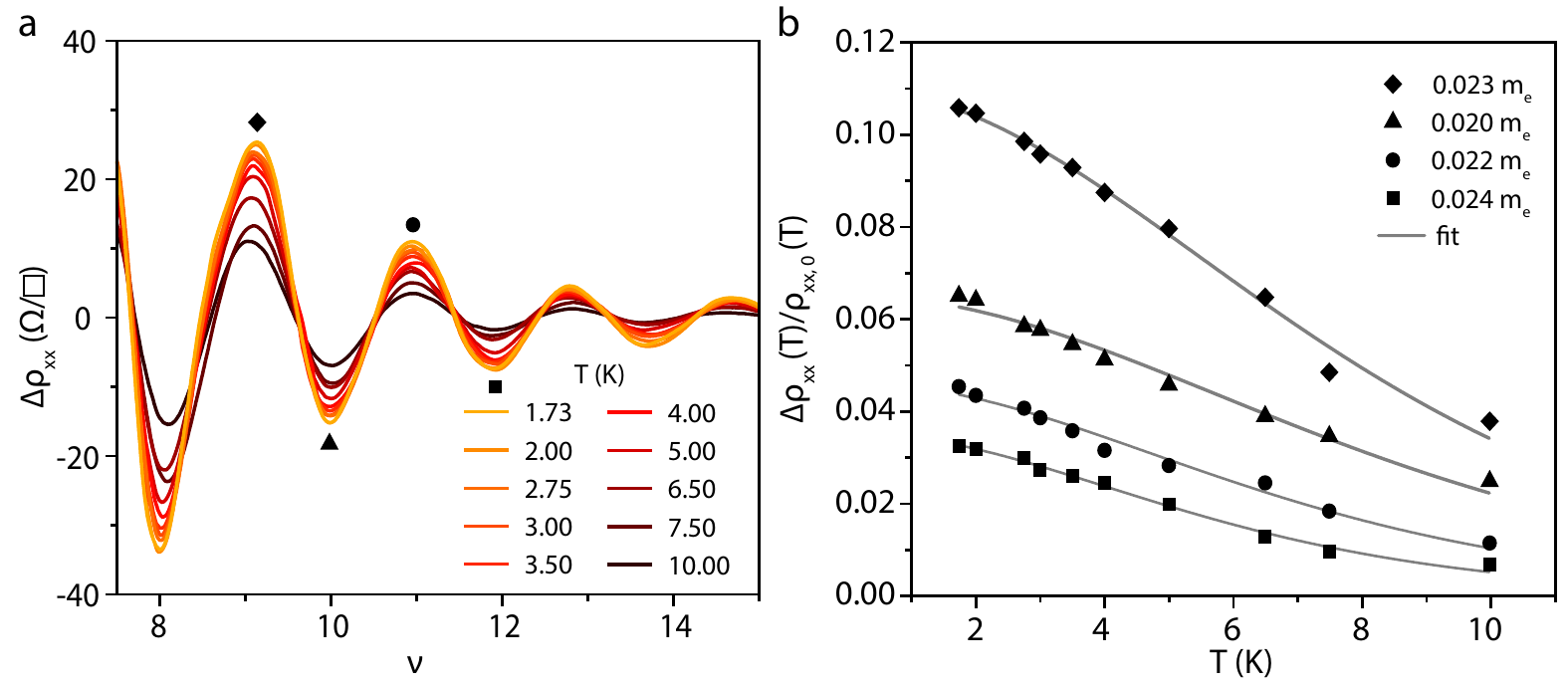}
	\caption{(a), Shubnikov-de Haas oscillation amplitude after polynomial background subtraction as a function of filling factor for temperatures $T=1.73-10~\text{K}$. The symbols denote points that are used to extract the effective mass. (b), Temperature dependence of the oscillation amplitude (symbols). The solid lines are fits to the data (using Eq. \ref{Eq.Meff}) in order to obtain the effective mass.}
	\label{SF.6}
\end{figure*}

To extract the effective mass of the electrons in the InSb 2DEG, the temperature dependence of the Shubnikov-de Haas (SdH) oscillation amplitude is measured in a Hall bar geometry. Figure \ref{SF.6}a shows the magnetoresistance oscillations after the subtraction of a polynomial background, $\Delta\rho_{\text{xx}}$, as a function of filling factor, $\nu$, for temperatures ranging from $T=1.73~\text{K}$ to $T=10~\text{K}$. At a fixed filling factor, the effective mass, $m^*$, can be obtained from a fit to the damping of the SdH oscillation amplitude with increasing temperature, using the expression
\begin{equation}
\frac{\Delta\rho_{\text{xx}}(T)}{\rho_{\text{xx,0}}(T)}\propto\frac{\alpha T}{\text{sinh}(\alpha T)}~, \label{Eq.Meff}
\end{equation}
where $\rho_{\text{xx,0}}(T)$ is the temperature-dependent low-field resistivity and $\alpha=\pi k_{\text{B}} m^{*}\nu/(\hbar^2 n)$. Figure \ref{SF.6}b shows such fits to the oscillation minima and maxima of $\nu=10$ and $\nu=12$, resulting in a mean effective mass of $m^*=(0.022 \pm 0.002)\cdot m_e$, with $m_e$ being the free electron mass. 

\newpage


\begin{thebibliography}{44}
\bibitem [{ref({\natexlab{b}})}]{refnotesoi}%
\BibitemOpen
\bibinfo {note} {The depicted shift of the Fermi surfaces assumes that the
	spin-orbit energy dominates over the Zeeman energy, which is the case for the
	measured JJs. See SI for a detailed discussion on this.}\BibitemShut {Stop}%
\end{thebibliography}
\end{document}